\documentclass[aps,preprint,superscriptaddress]{revtex4-2}
\usepackage{graphicx,color}
\usepackage{dcolumn}
\usepackage{bm}
\usepackage[utf8]{inputenc}
\usepackage[T1]{fontenc}
\usepackage{mathptmx}
\usepackage{etoolbox}
\usepackage{times}
\usepackage{amsfonts,amsmath,amssymb,amsbsy}
\usepackage[colorlinks=true, linkcolor=blue, citecolor=blue, urlcolor=blue]{hyperref}
\usepackage{float}
\usepackage{lettrine}
\def\D{{\textbf{D}}}
\def\H{{\textbf{H}}}
\def\h{{\textbf{h}}}
\def\m{{\textbf{m}}}
\def\p{{\textbf{p}}}
\def\F{{\textbf{F}}}
\def\G{{\textbf{G}}}
\def\v{{\textbf{v}}}
\let\mathbf=\boldsymbol

\def\section#1{\medskip\noindent\textbf{#1}\par}
 
\begin{document}

\title{\sffamily Elongated Skyrmion as Spin Torque Nano-Oscillator and Magnonic Waveguide}

\author{\sffamily Xue Liang}
\affiliation{\sffamily School of Science and Engineering, The Chinese University of Hong Kong, Shenzhen, Guangdong 518172, China}

\author{\sffamily Laichuan Shen}
\affiliation{\sffamily The Center for Advanced Quantum Studies and Department of Physics, Beijing Normal University, Beijing 100875, China}

\author{\sffamily Xiangjun Xing}
\affiliation{\sffamily School of Physics and Optoelectronic Engineering, Guangdong University of Technology, Guangzhou 510006, China}

\author{\sffamily Yan Zhou}
\email[E-mail:~]{zhouyan@cuhk.edu.cn}
\affiliation{\sffamily School of Science and Engineering, The Chinese University of Hong Kong, Shenzhen, Guangdong 518172, China}

{\sffamily
\begin{abstract}
\noindent
Spin torque nano-oscillator has been extensively studied both theoretically and experimentally in recent decades due to its potential applications in future microwave communication technology and neuromorphic computing. In this work, we present a skyrmion-based spin torque nano-oscillator driven by a spatially uniform direct current, where an elongated skyrmion is confined by two pinning sites. Different from other skyrmion-based oscillators that arise from the circular motion or the breathing mode of a skyrmion, the steady-state oscillatory motion is produced by the periodic deformation of the elongated skyrmion, which originates from the oscillation of its partial domain walls under the joint action of spin torques, the damping and the boundary effect. Micromagnetic simulations are performed to demonstrate the dependence of the oscillation frequency on the driving current, the damping constant, the magnetic parameters as well as the characteristics of pinning sites. This nonlinear response to a direct current turns out to be universal and can also appear in the case of elongated antiskyrmions, skyrmioniums and domain walls. Furthermore, the elongated skyrmion possesses a rectangle-like domain wall, which could also serve as a magnonic waveguide. These findings will enrich the design options for future skyrmion-based devices in the information technology.
\end{abstract}}

\date{\today}

\maketitle

\clearpage

\vbox{}
\section{\sffamily Introduction}

Magnetization precession is among the primary phenomena in micromagnetics and spintronics, which can be triggered by various external excitations, such as the applied field and the spin-polarized current~\cite{Slonczewski_JMMM1996,Berger_PRB1996,Kiselev_NATURE2003,Demidov_PR2017,Deac_NATPHY2008,Rippard_PRL2004}. The latter is intensively studied and has received much attention for potential applications in nano-oscillators, as it can effectively excite the oscillation of magnetization without the needs of external magnetic fields. The sustainable oscillation originates from the compensation between the current-induced spin transfer torques and the intrinsic damping, and therefore, the corresponding devices are called spin torque nano-oscillators (STNOs)~\cite{Kiselev_NATURE2003,Awad_NATPHY2017,Silva_JMMM2008,Slavin_IEEE2009,Zeng_NANOSCA2013,Chen_IEEE2016,Tarequzzaman_CP2019}. Such a nonlinear phenomenon focuses on the persistent precession of uniform magnetization~\cite{Kiselev_NATURE2003,Rippard_PRL2004,Tarequzzaman_CP2019,WuTH_PRB2021} and later extends to the magnetic vortex~\cite{Pribiag_NATPHY2007,LiC_JMMM2020,Kasai_PRL2006,Zeng_APL2021,Dussaux_NC2010,Jenkins_CP2021}, domain walls~\cite{Martinez_PRB2011,Voto_SCIREP2017,Toro_JAP2020}, skyrmions~\cite{ZhangSF_NJP2015,Garcia-Sanchez_NJP2015,Sisodia_PRB2019,XingR_PRA2020,YuG_APL2021,Chui_AIP2015,JinC_NJP2020,ShenLC_APL2019,ShenLC_APL2022}, etc. Meanwhile, the dynamics of magnetic oscillation produces spin wave, which is also known as magnon in the quantum context and carries spin angular momentum~\cite{KajiwaraY_NAT2010,ChumakAV_NP2015}. This collective excitation is similar to acoustic or optical waves, possessing fundamental wave properties, such as interference, diffraction, and polarization. 

On the other hand, magnetic skyrmions are topologically protected swirling spin textures, which have been experimentally observed in various magnetic systems, including ferromagnets~\cite{Muhlbauer_SCI2009,Yu_NATURE2010,Heinze_NATPHY2011,JiangW_SCI2015,Woo_NATMAT2016,Maccariello_NATTECH2018,Peng_NATTECH2020}, ferrimagnets~\cite{Woo_NATCOM2018}, and multiferroic materials~\cite{Seki_SCI2012,WangY_NATCOM2020} and whose polar counterpart has been observed in ferroelectric materials~\cite{Nahas_NATCOM2015,Das_NATURE2019}. Due to their unique static properties and rich dynamics, they are widely considered to be the information carriers for future ultrahigh-density, energy-efficient spintronics devices.~\cite{Fert_NATTECH2013,Sampaio_NATTECH2013,Tomasello_SCIREP2014,ZhangXC_SCIPER9400,Zhou_NSR2019,Gobel_PR2021,ZhangXC_JPCM2020}. Among them, skyrmion-based STNO is one of the most important applications, and several schemes have been proposed to demonstrate the high frequency microwave generation~\cite{ZhangSF_NJP2015,Garcia-Sanchez_NJP2015,Sisodia_PRB2019,XingR_PRA2020,YuG_APL2021,Chui_AIP2015,JinC_NJP2020,ShenLC_APL2019}. To date, there are three main mechanisms: 1) the gyrotropic motion of skyrmions in a nanodisk induced by spin torques and the repulsive force from the boundary~\cite{ZhangSF_NJP2015,Garcia-Sanchez_NJP2015,Sisodia_PRB2019,ShenLC_APL2019}; 2) the breathing mode of skyrmions characterized by periodic changes in size~\cite{YuG_APL2021,Chui_AIP2015,Lonsky_PRB2020,KimJV_PRB2014}; 3) the periodic deformation of a skyrmion driven by a spatial-dependent spin current~\cite{XingR_PRA2020}. 

Compared with the circular motion and breathing mode of skyrmions, the oscillation produced by the third mechanism is much less studied. In this work, we propose an optional scheme to present the skyrmion oscillation induced by the periodic deformation, where an elongated skyrmion is confined in a nanotrack with two pinning sites. Within a certain range of the driving current density, the magnetic potential well induced by the pinning regions and the boundary as well as the competition between the spin torques and inherent damping yield a sustained oscillation of magnetization. The frequency can reach a few GHz that is higher than most of other ferromagnetic skyrmion-based oscillators (around a few hundred MHz). The oscillation is also caused by the deformation of a skyrmion, but the driving current is spatially uniform, which is easily accessible compared to the method reported in other work~\cite{XingR_PRA2020}. Furthermore, both theoretical and experimental results have shown that a domain wall naturally acts as a magnonic waveguide, where spin waves can be excited easily and propagate with low dissipation~\cite{Garcia-Sanchez_PRL2015,XingXJ_NPG2016,Wagner_NATTECH2016,Albisetti_CP2018}. In our model, the elongated skyrmion involves a rectangle-like domain wall so that the spin wave can propagate in this channel. From the application point of view, both STNOs and magnonic waveguides are promising building blocks for the next-generation information processing and storage devices, as well as the pattern recognition and classification in neuromorphic computing~\cite{Torrejon_NATURE2017,Romera_NATURE2018,Zahedinejad_NATTECH2020,Markovic_APL2019,PappA_NC2021,KozhevnikovA_APL2015}.
%

\vbox{}
\section{\sffamily Results and Discussion}

\begin{figure*}[t!]
\centerline{\includegraphics[width=0.8\textwidth]{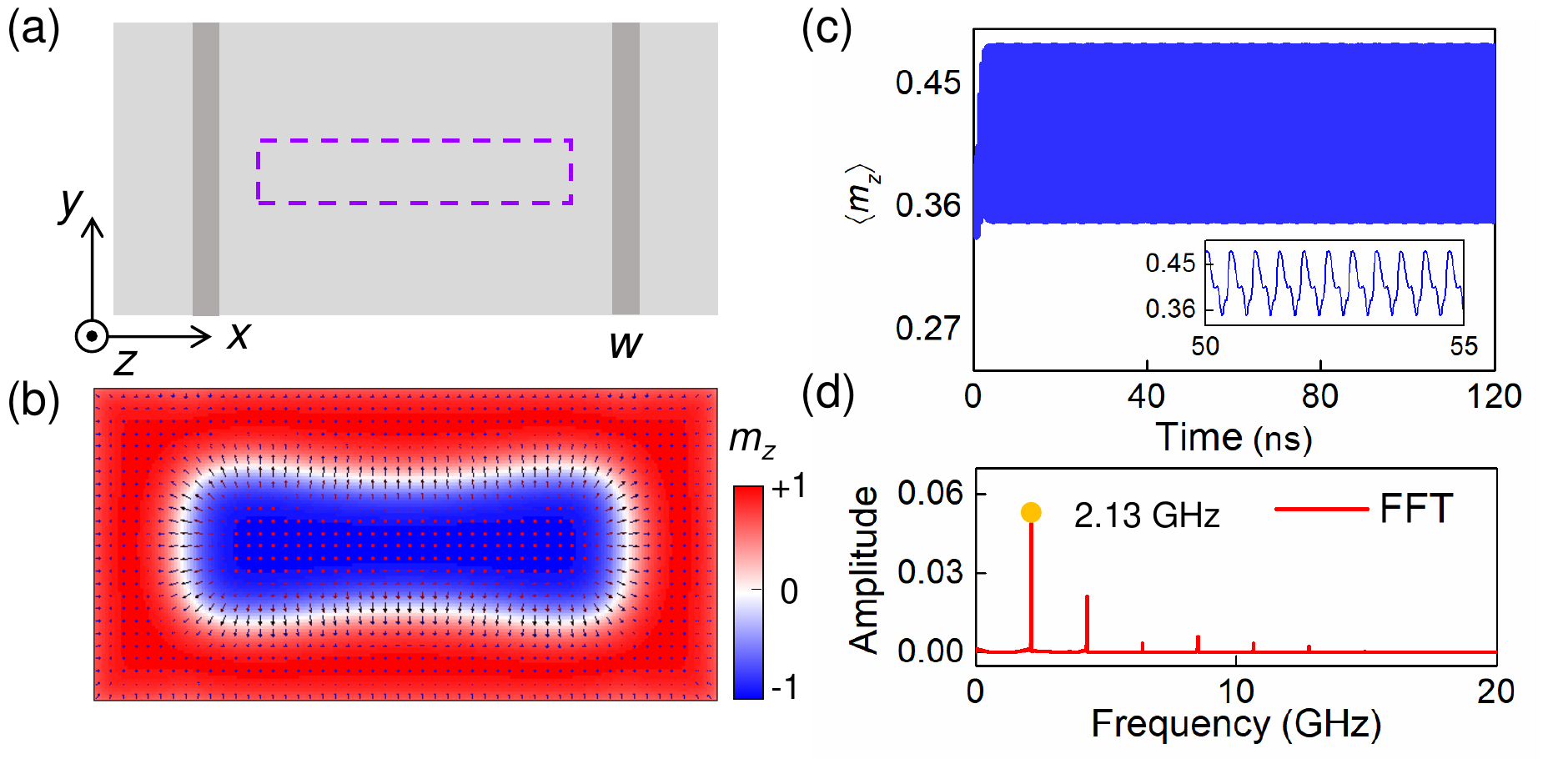}}
\caption{
\textbf{Schematic diagram of an elongated-skyrmion-based oscillator and the response of magnetization to a spatially uniform direct current.}
(a) Schematic diagram of the ferromagnetic nanotrack utilized for stabilizing a pinned elongated skyrmion. $w$ is the width of pinning sites.
(b) The spin configuration of a relaxed elongated skyrmion, where colors represent the out-of-plane component of the local unit magnetization $m_z$ as indicated by the color bar. 
(c) Spatially averaged magnetization $\langle m_z \rangle$ of the entire film oscillates periodically when the driving current density is 6 MA$\cdot$cm$^{-2}$. Inset shows the enlarged oscillations in 5.0 ns.
(d) Frequency spectra of this oscillation is obtained from the fast Fourier transform (FFT) of $\langle m_z \rangle$ in (c).
}
\label{FIG1}
\end{figure*}

\noindent
\textbf{Model and initial state.}
Here, we consider a ferromagnetic nanotrack grown on a heavy metal that provides sufficiently strong Dzyaloshinskii-Moriya interaction (DMI)~\cite{DMID1958,DMIM1960} to stabilize chiral skyrmion-like structures and allows the spin Hall effect when an electric current flows through it due to the strong spin-orbit interaction. To obtain an elongated skyrmion, two artificial pinning sites are placed in the nanotrack, which are constructed by a local perpendicular magnetic anisotropy (PMA) reduction and indicated by the dark gray region of width $w$ in Fig.~\ref{FIG1}(a). The PMA constant in pinning sites is $K_{\text{p}}=0.55$ MJ$\cdot$m$^{-3}$  and the width is $w=10$ nm in the following simulation, unless specified otherwise. Such a modification of PMA considered in this device is due to the fact that the local PMA reduction can generate an energy well to attract or trap skyrmions and act as a pinning site~\cite{ZhangX_SCIR2015K,LiangX_PRB2019,Castell_NANOSCALE2019}. Recent theoretical work has also reported the formation of square-shaped skyrmions in a magnetic thin film with orthogonal defect lines formed by the reduction of PMA~\cite{Xichao_CP2021}. In experiments, this type of pinning sites can be created by locally applying additional sputtered layers~\cite{Ohara_NANOLETT2021} or by using ion irradiation~\cite{JugeR_NLETT2021}.

We begin with the nucleation of an elongated skyrmion. At the first step, a single spin-polarized current pulse is perpendicularly injected into the central region indicated by the dashed box in Fig.~\ref{FIG1}(a) to overcoming the topological stability barrier. The current density and the duration of this pulse are 300 MA$\cdot$cm$^{-2}$ and 0.2 ns, respectively. Then, the current is switched off, and a stable elongated skyrmion confined in the region between two pinning sites is obtained after a sufficiently long-time relaxation (see Supplementary Note 1 for more details). This current-induced generation of skyrmions is one of the most efficient schemes for achieving the reproducible nucleation of individual skyrmions at a given position~\cite{Sampaio_NATTECH2013}, which has also been experimentally realized via vertical current injection through a scanning tunneling microscope tip~\cite{Romming_SCIENCE2013}, 3D conducting path~\cite{Yang_AM2021} or magnetic tunnel junctions~\cite{Penthorn_PRL2019}. The rapid progress in this field provides promising future for the experimental realization of the proposed device. Figure~\ref{FIG1}(b) shows the spin configuration of a relaxed elongated skyrmion. Indeed, the formation of this non-circular spin texture can be seen as a circle-shaped skyrmion elongated by two forces from the pinning sites since the region with reduced PMA tends to attract a domain wall or a skyrmion as mentioned before. Supplementary Note 2 provides details of the conversion between a circle-shaped ordinary skyrmion and an elongated skyrmion. 

\noindent
\textbf{Oscillations of a domain wall and an elongated skyrmion.}
Next, when the driving current is applied to the heavy metal along $x$ axis, the elongated skyrmion will deform periodically and generate a stable spin oscillation that is detected by the magnetoresistance effect through a magnetic tunnel junction. To quantify this dynamical behavior, we plot the temporal evolution of $\langle m_z \rangle$ as shown in Fig.~\ref{FIG1}(c), where $\langle\cdots\rangle$ represents the spatial average of a component of magnetization over the entire nanotrack, and calculate the oscillation frequency by using the fast Fourier transform (FFT). From the frequency spectra in Fig.~\ref{FIG1}(d), it is noted that some higher-order harmonics are also excited, in addition to the fundamental frequency. Therefore, the profile of $\langle m_z \rangle$ is not a pure sine-like curve, as illustrated in Fig.~\ref{FIG1}(c).   
%

\begin{figure*}[t!]
\centerline{\includegraphics[width=0.82\textwidth]{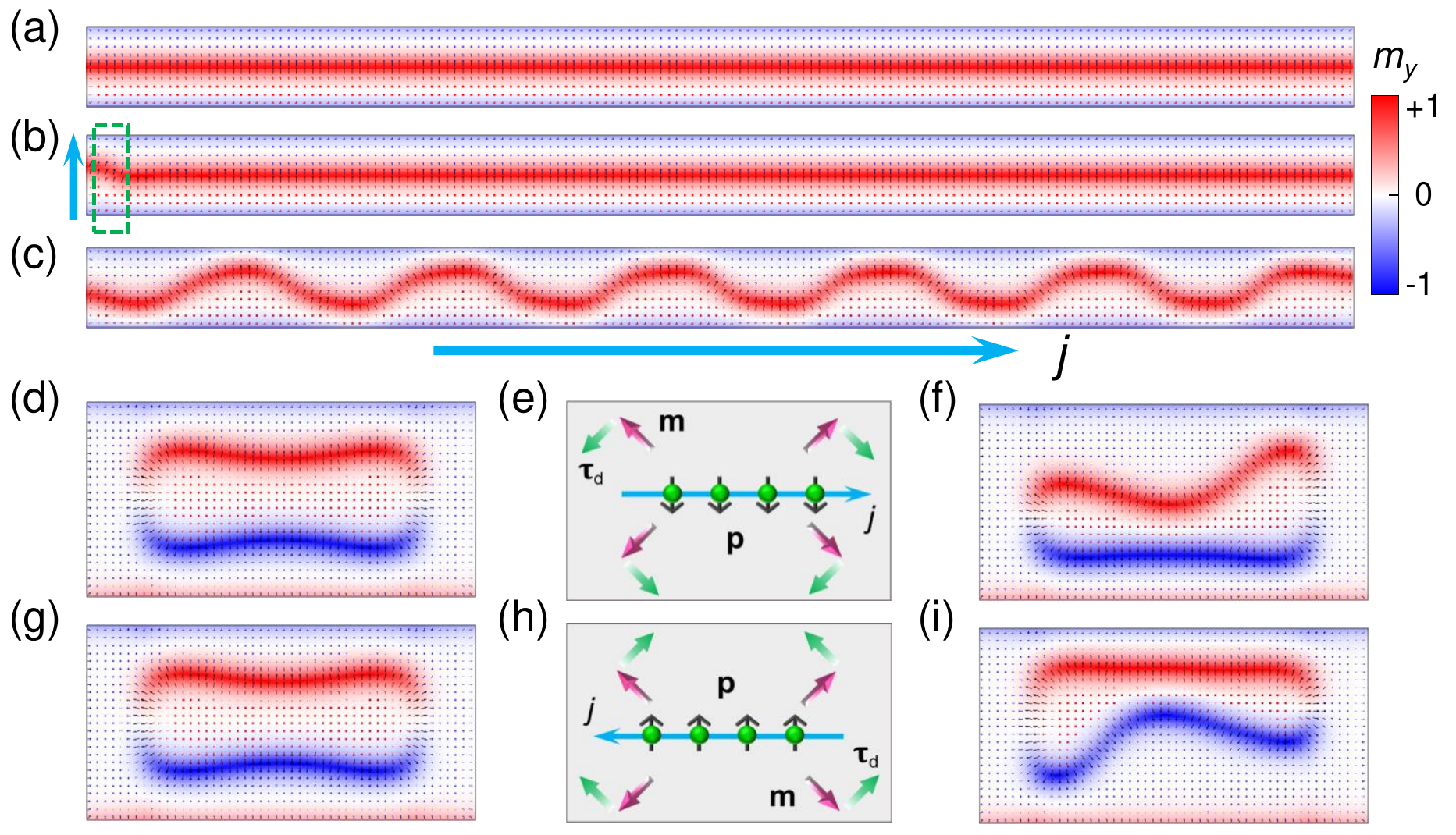}}
\caption{
\textbf{Current-induced oscillation of a domain wall and an elongated skyrmion.
}%
Snapshots of the magnetization of a domain wall at (a) equilibrium, (b) initialization and (c) oscillation. The color bar represents the $y$ component of the magnetization $m_{y}$.
The skyrmion oscillation driven by the spin current with the polarization (d-f) $\p=-\hat{y}$ and (g-i) $\p=+\hat{y}$. Here, (d) and (g) represent the initial states, (e) and (h) are the schematic diagram of the local magnetization $\m$ (denoted by the pink arrows) and the damping-like torque ${\boldsymbol{\tau}}_{\text{d}}$ (denoted by the green arrows) induced by a spin-polarized current $j$ flowing along the blue arrow, (f) and (i) depict the snapshot of the magnetization at a selected time during the oscillation.
}
\label{FIG2}
\end{figure*}

To understand the skyrmion oscillation, we first consider the case where a single domain wall is located at the center of a nanotrack as depicted in Fig.~\ref{FIG2}(a). When the spin current is applied parallel to the domain wall (i.e., the corresponding unit polarization vector is $\p=\pm\hat{y}$, since $\p$ should be perpendicular to the current flow in the case of spin Hall effect), the domain wall can not be driven and remain unchanged. However, if there is a perturbation to force the partial domain wall to deform and deviate from the equilibrium at the initial state as shown in Fig.~\ref{FIG2}(b), the domain wall will create a sustained oscillation driven by the spin current along the $x$ axis, similar to an ordinary rope [see Fig.~\ref{FIG2}(c) and Supplementary Movie 1]. Here, the oscillation is attributed to the competition between the spin torque, the intrinsic damping and the boundary barrier. Besides, such a perturbation can be realized by locally applying an external field or a driving current along $y$ axis. However, it should be noted that the oscillation occurs only when the spin polarization of the driving current is opposite to the orientation of the $y$ component of magnetization in the domain wall, e.g., $\p$ is equal to $-\hat{y}$ for the case shown in Fig.~\ref{FIG2}(a).  Otherwise, the domain wall will restore to equilibrium and keep steady. As a result, the elongated skyrmion is considered as an oscillation source in this work with two significant advantages: 1) it possesses natural domain walls with curvature so that no artificial initialization is required to produce a perturbation; 2) Whether the spin polarization is $+\hat{y}$ or $-\hat{y}$, the oscillation is detected, since the elongated skyrmion has two types of domain walls, corresponding to the reversal of the magnetization from $+\hat{z}$ to $-\hat{z}$ and from $-\hat{z}$ to $+\hat{z}$, respectively. Figures~\ref{FIG2}(d-i) show the distribution of the in-plane magnetization component and the schematic diagram of the current-induced damping-like torque ${\boldsymbol{\tau}}_{\text{d}}$ acting on the local magnetization with the unit vector $\m$. It is seen that, driven by the spin-polarized current with $\p=-\hat{y}$ in Figs.~\ref{FIG2}(d-f) and Supplementary Movie 2, the top domain wall (i.e., the red stripe) periodically oscillates, while the bottom one just has a slight drift. Considering the fact that the local magnetization tends to align in the direction of polarization in the presence of spin torque, the magnetic moments in the red stripe are more fragile to reverse than that in blue one and altering the polarization direction of the driving current can switch the oscillating section of this elongated skyrmion as shown in Figs.~\ref{FIG2}(g-i). 
%

\begin{figure*}[t]
\centerline{\includegraphics[width=0.85\textwidth]{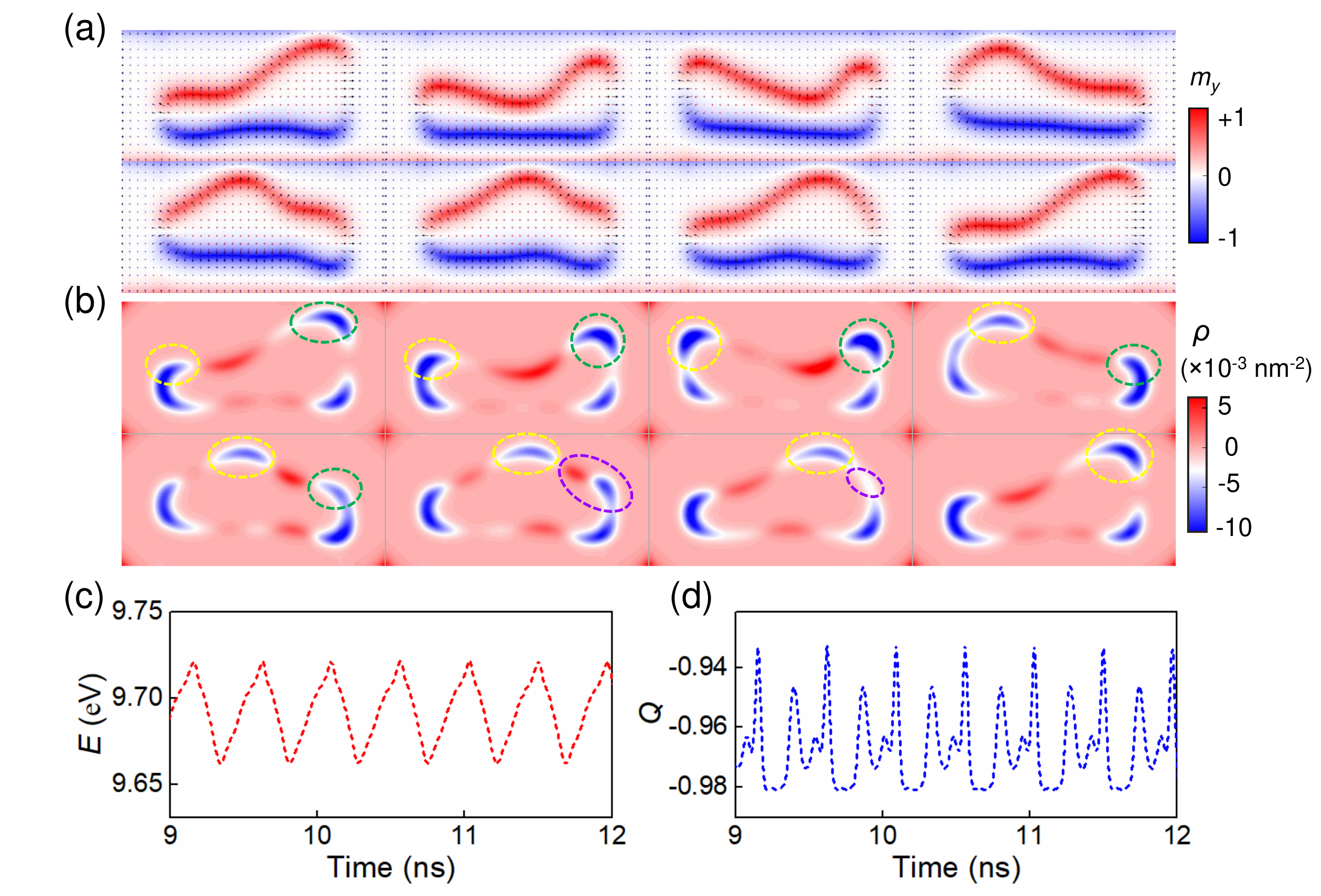}}
\caption{
\textbf{Periodic deformation of an elongated skyrmion during the oscillation.
}%
Snapshots of (a) the magnetization and (b) the corresponding spatial distributions of the topological charge density at different simulation time in a period. The color bar in (a) represents the $y$ component of the magnetization $m_{y}$, while it describes the topological charge density $\rho$ in (b).
Variations of (c) the total energy $E$ and (d) the net topological number $Q$ during the oscillation.
}
\label{FIG3}
\end{figure*}

\begin{figure*}[t!]
\centerline{\includegraphics[width=0.78\textwidth]{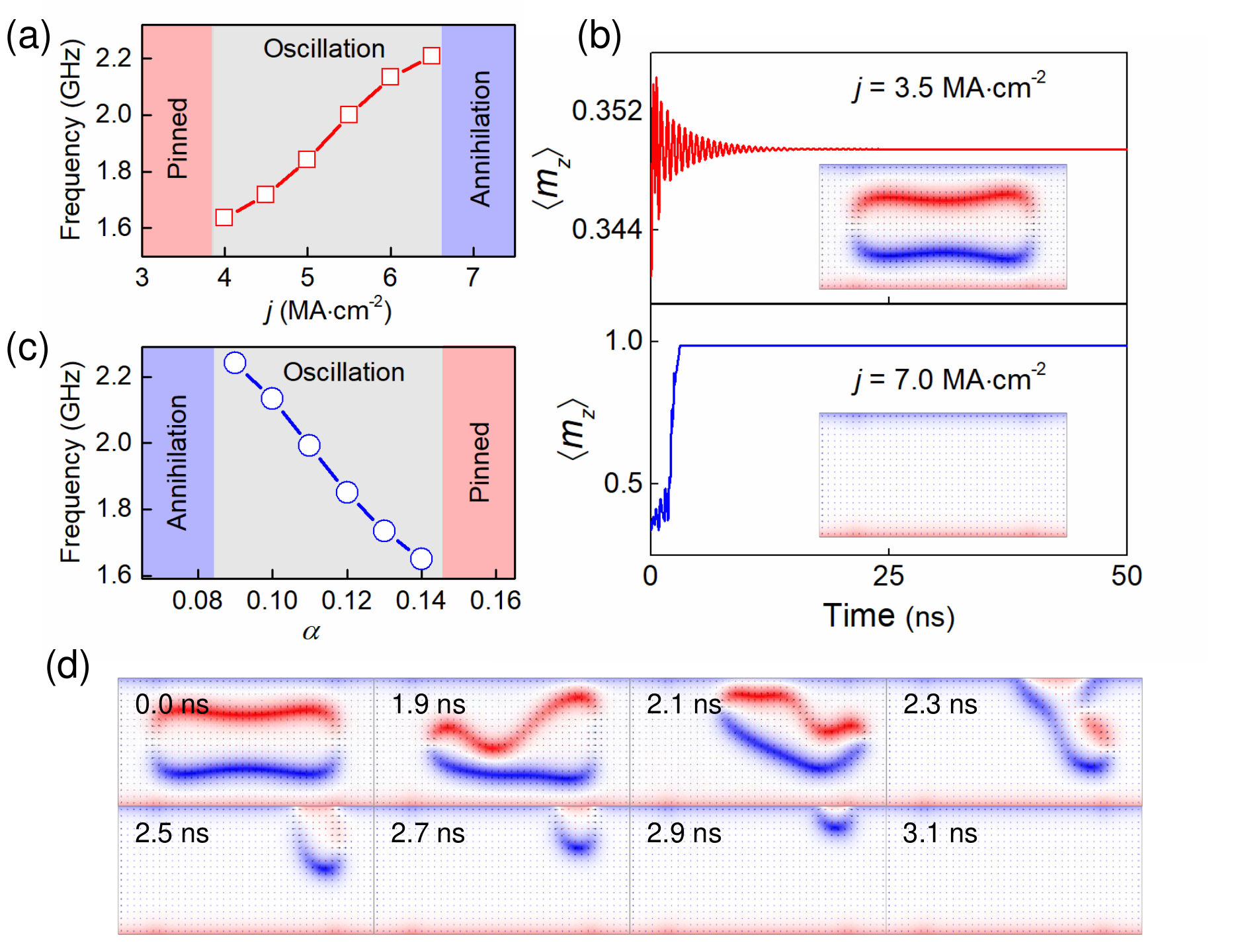}}
\caption{
\textbf{Effects of the current density and the damping on the oscillation of an elongated skyrmion.
}%
(a) Frequency of the considered oscillation as a function of the driving current density $j$, where the pink and purple regions denote the pinned phase and annihilation phase, respectively.  
(b) Time-dependent of the magnetization $\langle m_z \rangle$ in the cases of pinned and annihilation. Insets are the final stable magnetization configurations.
(c) Oscillation frequency as a function of the damping coefficient $\alpha$.
(d) Snapshots of the dynamical annihilation process of an elongated skyrmion. 
}
\label{FIG4}
\end{figure*}

In addition, eight snapshot images of the time-dependent spin configuration and the corresponding spatial distribution of the topological charge density defined as $\rho=\m\cdot({\partial}_x\m\times{\partial}_y\m)$ within one period are shown in Figs.~\ref{FIG3}(a) and (b). We use the method proposed by S.-Z. Lin ~\cite{LinSZ_PRB2016} to further describe the magnetization dynamics, where the overall deformation of a spin configuration is reflected in the motion of half-skyrmions. In Fig.~\ref{FIG3}(b), the element marked by the yellow dashed circle is formed by the accumulation of negative topological charges, which is regarded as a half-skyrmion with negative fractional topological number $Q=({1}/{4\pi})\int {\rho}\mathrm{d}x\mathrm{d}y$. In the left pinning site, this half-skyrmion moves upward along the border of the pinning region, and gradually separates from the pinned part at the bottom. Due to the repulsive force from the upper boundary of the nanotrack, it cannot continue to move upwards, but moves to the right pinning site under the action of a spin-polarized current. Now, we turn to tracking the upper half-skyrmion marked by the green dashed circle in the right pinning site. Since the force provided by the driving current is not enough to overcome the potential barrier of the pinning region, the half-skyrmion is confined in this area and moves downward to merge with the lower part of negative topological charge. Afterwards, it moves up analogous to the left marked half-skyrmion. However, due to the approach of the half-skyrmion marked by yellow circle, positive topological charges are induced between the two negative topological charge accumulation regions, corresponding to the bend of the domain wall string between two protrusions as shown in Fig.~\ref{FIG3}(a). Consequently, the upward motion of the half-skyrmion marked by the green circle will be suppressed, and it tends to compensate with the positive charges as indicated by the purple dashed circle. The half-skyrmion from the left side finally reaches the right pinning site and becomes the half-skyrmion marked by the green circle for the next cycle. This process occurs repeatedly and produces a stable spin oscillation without any decay. Figures~\ref{FIG3}(c) and (d) exhibit the total energy and the net topological number as a function of the time during the oscillation. It is seen that, the oscillating variation in both the energy and the topological number also implies the periodic deformation of this elongated skyrmion. 
%

\begin{figure*}[t]
\centerline{\includegraphics[width=0.85\textwidth]{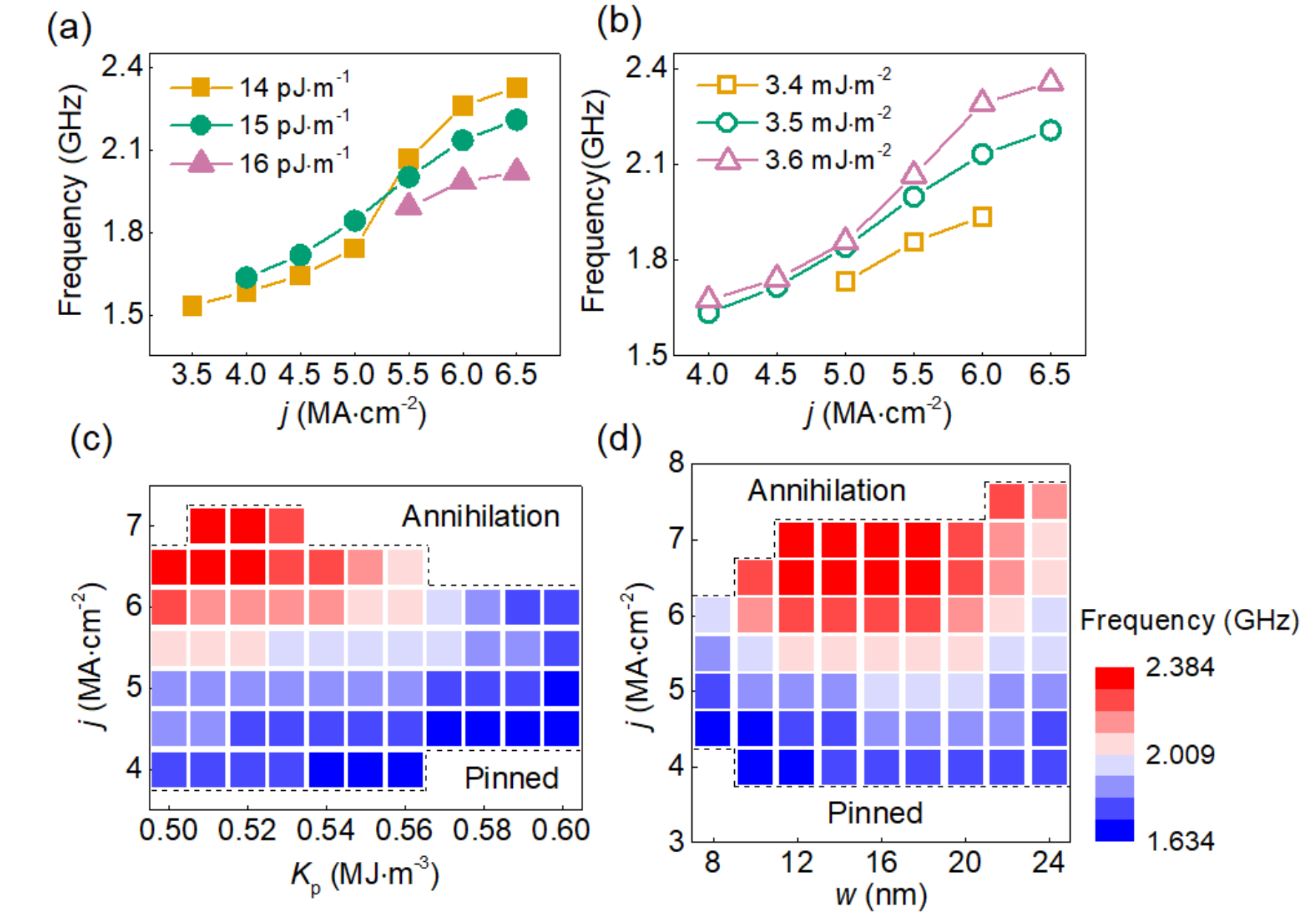}}
\caption{
\textbf{Parametric dependence of oscillation frequency.}
The oscillation frequency as a function of the driving current density $j$ for different exchange stiffness $A$ and Dzyaloshinskii-Moriya interaction constant $D$. Here, $D$ is fixed at 3.5 mJ$\cdot$m$^{-2}$ in (a), and $A=15$ pJ$\cdot$m$^{-1}$ in (b). The yellow filled squares, green filled circles and pink filled triangles in (a) denote the case of $A=14, 15$ and 16 pJ$\cdot$m$^{-1}$, respectively. The yellow squares, green circles and pink triangles in (b) represent the case of $D=3.4, 3.5$ and 3.6 mJ$\cdot$m$^{-2}$, respectively.
Phase diagrams of the oscillation versus the current density $j$ and (c) the perpendicular magnetic anisotropy value $K_{\text{p}}$ or (d) the width $w$ of the pinning sites, where $A=15$ pJ$\cdot$m$^{-1}$, $D=3.5$ mJ$\cdot$m$^{-2}$ and the color bar represents the oscillation frequency.
}
\label{FIG5}
\end{figure*}

Figure~\ref{FIG4}(a) demonstrates the relationship between the oscillation frequency and the driving current density. It is found that there are three distinct regions in the whole working window, as illustrated by different color backgrounds. As shown in Fig.~\ref{FIG4}(b), when the driving current density $j$ is smaller than the first critical value $j_1$, the elongated skyrmion is completely pinned, where the initial $\langle m_z \rangle$ oscillation gradually decays to 0, indicating that the driving force is insufficient to overcome the intrinsic force due to the topological protection and the damping force. However, if the current density is larger than the second critical value $j_2$, the elongated skyrmion will depin from the pinning sites and be annihilated around the upper boundary of the nanotrack [see Fig.~\ref{FIG4}(d)], which is similar to the motion of a skyrmion in a homogeneous nanotrack driven by a large current. In the range of $j_1<j<j_2$, the steady skyrmion oscillation occurs, and the frequency increases with increasing $j$. The dependence of oscillation on the damping coefficient $\alpha$ is also shown in Fig.~\ref{FIG4}(c). There are still three phases, but, on the contrary, the pinned and annihilation states occur in the high-damping and low-damping areas, respectively, and the frequency is inversely proportional to $\alpha$ in the oscillation region. For a rigid topological object driven by a spin-polarized current, the Thiele equation is given by~\cite{Thiele_PRL1973}
\begin{equation}
\begin{split}
\label{eq:Thiele}
\G\times\v+\alpha{\D}\cdot\v=\F_{j}+\F_{r},
\end{split}
\end{equation}
where the first term represents the Magnus force with $\G=(0,0,G)$ and $G=4\pi{Q}$, $\D$ is the dissipative tensor with $D_{ij}=\delta_{ij}\mathcal{D}$, $\F_{j}$ denotes the current-induced driving force proportional to the current density $j$ and $\F_{r}$ is the repulsive force imposed by the boundary. If this driving current is applied along $x$ axis with $\p=-\hat{y}$ and $\F_{r}=F_r\hat{y}$, we obtain the following velocity relationship:
\begin{equation}
\begin{split}
\label{eq:v}
\left(
\begin{array}{c}
v_{x}\\
v_{y}
\end{array}
\right)=\frac{1}{\alpha^2\mathcal{D}^2+G^2}\left(
\begin{array}{c}
\alpha\mathcal{D}F_{j}+GF_{r}\\
-GF_{j}+\alpha\mathcal{D}F_{r}
\end{array}
\right).
\end{split}
\end{equation}
When the topological body moves along the upper boundary of the nanotrack, the Magnus force is balanced by the repulsive force from the edge, $v_y=0$ and $v_x=F_j/(\alpha \mathcal{D})$. It should be noted that the velocity is proportional to the driving current density and increases as the damping constant decreases. From the perspective of qualitative analysis, a fast motion of the marked half-skyrmions in Fig.~\ref{FIG3}(b) requires a large current density and low damping, which can reduce the motion period and hence increase the oscillation frequency. One can understand why the oscillation frequency increases with increasing $j$ and is inversely proportional to the damping constant.

Here, we continue to investigate the skyrmion oscillation driven by a spin-polarized current as a function of the magnetic parameters. It should be mentioned that the Heisenberg exchange interaction tends to force the magnetic moments of adjacent atoms to align with one another in parallel, while the DMI prefers them to form a non-collinear structure. Therefore, a weak Heisenberg exchange interaction and a strong DMI allow the elongated skyrmion to deform easily and then produce a oscillatory motion of magnetization as shown in Figs.~\ref{FIG5}(a) and (b). Meanwhile, the total energy versus the exchange stiffness $A$ and the DMI constant $D$ as well as the spin configurations with different magnetic parameters are also given in Supplementary Figure 5, suggesting that the system with a small $A$ or a large $D$ has a lower energy and is more likely to be excited to oscillate. Figures~\ref{FIG5}(c) and (d) show the effect of the characteristics of pinning sites on the oscillation. It is noted that for each of driving current density $j$, the increase in $K_{\text{p}}$ yields a reduction in the oscillation frequency, and the working window of $j$ is wider in low-PMA region, which is attributed to the strong pinning force caused by the large PMA difference between the pinning site and the clean region. On the other hand, the response of this oscillation to the width of pinning sites is non-monotonic, and there is an optimal size that corresponds to the maximum frequency. For a narrow pinning region, the strength of the pinning force is not strong enough to protect the skyrmion from annihilation. However, if the width $w$ is too large, most part of the skyrmion will be pinned, thus suppressing its deformation and oscillation. Consequently, to obtain a high frequency of the proposed skyrmion oscillator, the pinning site with low PMA and an appropriate optimal width $w$ should be chosen. In addition, it should be mentioned that this oscillation can also occur in the case of antiskyrmions, skyrmioniums (see Supplementary Movies 3 and 4).
%

\begin{figure}[t]
\centerline{\includegraphics[width=0.6\textwidth]{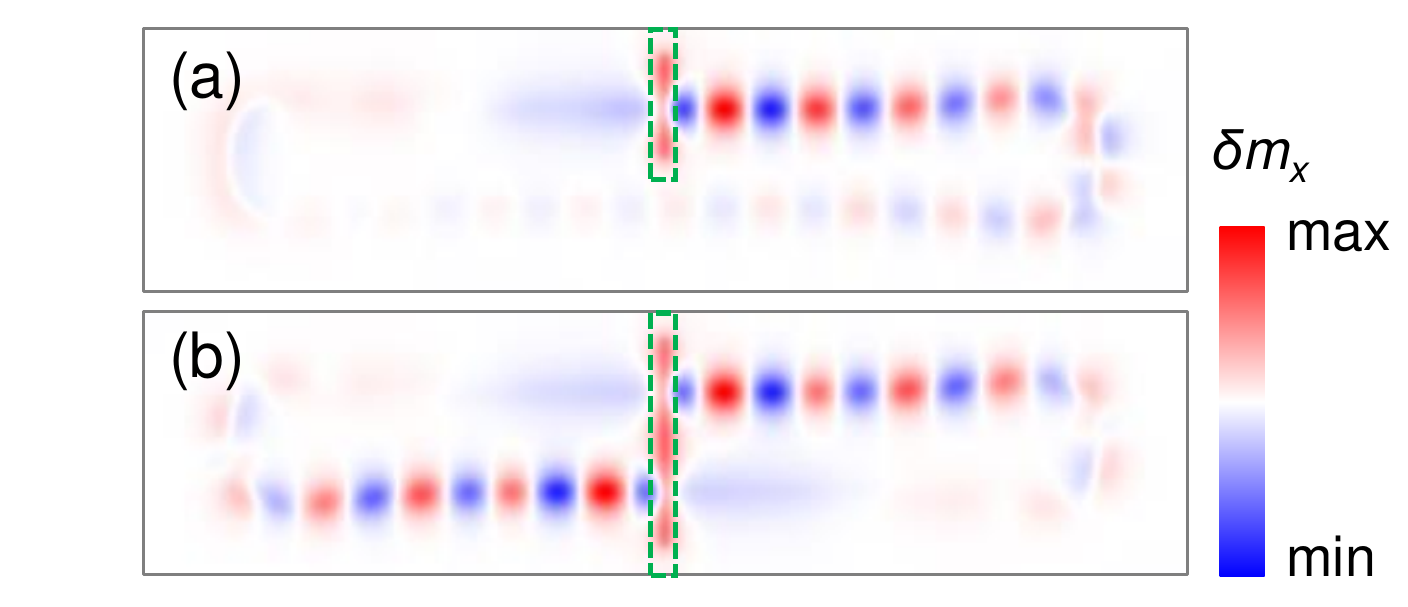}}
\caption{
\textbf{Spin-wave propagation patterns along the rectangle-like domain wall of an elongated skyrmion. }
Spatial profiles of the fluctuation in the $x$ component of magnetization $\delta m_x$ excited by a sinusoidal magnetic field $\h=h_0$sin$(2\pi ft)\hat{x}$, where $h_0$ and $f$ are the amplitude and the frequency of this excitation field, respectively. In our simulations, $h_0 = 20$ mT, $f = 20 $ GHz, the spatial area of the excitation field is (a) $4\times 50$ nm$^2$ and (b) $4\times 100$ nm$^2$. The color bar denotes the value of the fluctuation in the $x$ component of magnetization $\delta m_x$
}
\label{FIG6}
\end{figure}

\begin{figure}[t]
\centerline{\includegraphics[width=0.6\textwidth]{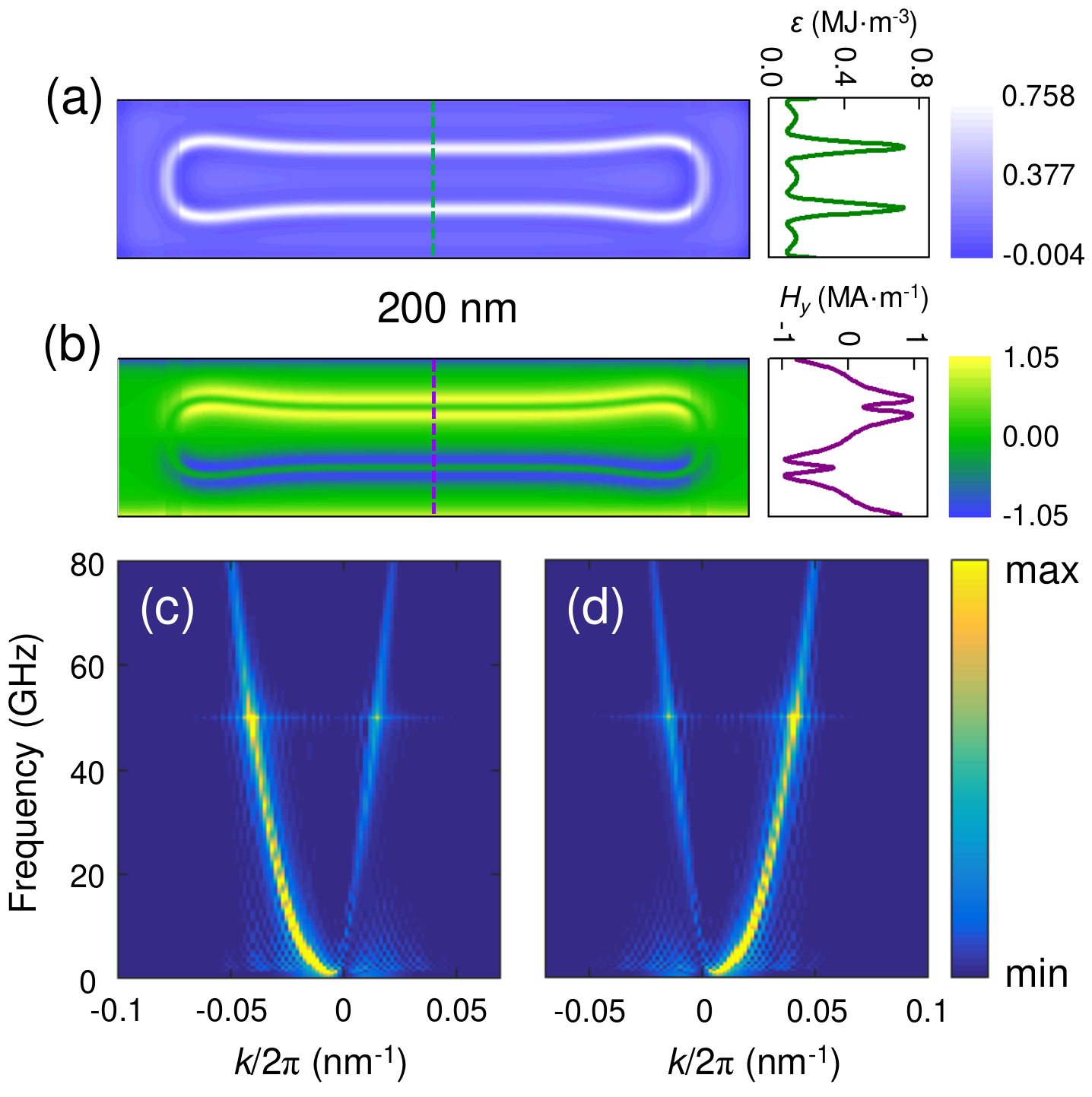}}
\caption{
\textbf{Properties of an elongated skyrmion acting as a magnonic waveguide.
}%
Maps of (a) the total energy density $\varepsilon$ and (b) the effective field $H^y_{\text{eff}}$ for the system with an elongated skyrmion. The insets at the right show the corresponding $\varepsilon$ and $H^y_{\text{eff}}$ along the central vertical line, $x=200$ nm. Dispersion relations for (c) the lower branch $y=30$ nm and (d) the upper branch $y=70$ nm of the elongated skyrmion. These three color bars denote the total energy density, the effective field and the fast Fourier transform (FFT) amplitude, respectively.
}
\label{FIG7}
\end{figure}

\noindent
\textbf{An elongated skyrmion as a magnonic waveguide.}
Furthermore, such an elongated skyrmion is also an ideal magnonic waveguide, in which the spin wave travels along the rectangle-like domain wall. For the spin-wave excitation, the most commonly used approach in experiments is to apply an inductive antenna, and the magnetization precession in magnetic materials is excited via the alternating Oersted field induced around the antenna~\cite{ChumakAV_NP2015,Chumak_APL2012,Demidov_APL2009}. The schematics of the proposed device with peripheral circuits is provided in Supplementary Note 3 and Supplementary Figure 6. In our simulation, we apply a local alternating magnetic field to excite spin waves. Note that the driving current previously used for the skyrmion oscillation should be removed in this section to study the spin wave propagation in an elongated skyrmion. We calculate the fluctuation in $x$ component of the magnetization $\delta m_x=m_x(t)-m_x(0)$ to demonstrate the spectra of spin waves in real-space-time domain intuitively, where $m_x(t)$ and $m_x(0)$ are the instant and the initial magnetization, respectively. First, the excitation field is applied to an area of $4\times 50$ nm$^2$, i.e., the top half of the nanotrack indicated by the dashed box in Fig.~\ref{FIG6}(a). It is seen that the excited spin waves propagating in two directions are nonreciprocal due to the presence of DMI~\cite{XingXJ_NPG2016,Garcia-Sanchez_PRL2015} and spread successfully from the upper branch to the lower branch along the domain wall. Extending the antenna to cover the nanotrack width [see Fig.~\ref{FIG6}(b)], we find that the spin waves generated on the two branches of the elongated skyrmion are asymmetrical. 

Figure~\ref{FIG7}(a) shows the distribution of the energy density, where the white channel corresponds to the rectangle-like domain wall of an elongated skyrmion with the maximum energy density. Excited by a low-frequency microwave stimuli, the magnetic moments in this region are easier to precess than that outside of the domain wall, thus confining spin waves in this narrow channel and forming a waveguide. In addition, considering that the directions of the magnetization or the effective field in the upper and lower branches are different [see Fig.~\ref{FIG1}(b) and Fig.~\ref{FIG7}(b)], one can understand why the propagation of excited spin waves is asymmetrical in Fig.~\ref{FIG6}(b). To further obtain the whole picture of the characteristics of wave propagation, the dispersion relations along two branches are also calculated, as shown in Figs.~\ref{FIG7}(c) and (d). The propagation of spin waves in such a ring-shaped waveguide will provide alternatives for designing of spintronics devices in future data processing and computing. Taking the logic gate as an example, we propose a device design to demonstrate its performance in Supplementary Note 4, which realizes the basic operations of logic AND, OR and XOR functions.
%

\vbox{}
\section{\sffamily Conclusions}

In this work, we demonstrate the current-induced oscillation of an elongated skyrmion on a nanotrack with two pinning sites. Combining the Thiele equation and the motion of half-skyrmions, the stable oscillation caused by the periodic deformation of an elongated skyrmion is described in detail. We show the effect of the driving current density, the damping coefficient, the magnetic properties and two primary parameters of the pinning region on the performance of the skyrmion oscillator. The working window of this oscillation is also determined. Furthermore, the waveguide properties of such an elongated skyrmion are corroborated through micromagnetic simulations. 

Our results reveal an alternative concept of  spin-torque oscillator based on a localized magnetic soliton, i.e., the elongated skyrmion. The resulting oscillation exhibits a nonlinear response to the direct current input, which is promising to be used in bio-inspired hardware. The utilization of nanoscale spintronic oscillators for high-performance neuromorphic computing has already been demonstrated in experiments~\cite{Torrejon_NATURE2017,Romera_NATURE2018,Zahedinejad_NATTECH2020}. For skyrmionics, from the experimental realization perspective, the discovery of ultra-thin magnetic materials with strong DMI for stabilizing isolated skyrmions and the development of nanofabrication techniques, would rapidly accelerate the skyrmion-based device optimization and its applications in future digital information technology.  Our work provides guidelines for the development and future experimental exploration of skyrmion-based oscillators.
%

\vbox{}
\section{\sffamily Methods}

\noindent
\textbf{Micromagnetic simulations.}
In this work, we use the public code project object-oriented micromagnetic framework (OOMMF)~\cite{OOMMF} to study the dynamics of magnetization based on the Landau-Lifshitz-Gilbert equation,
\begin{equation}
\label{eq:LLG}
\frac{d\m}{dt}=-\gamma\m\times\H_{\text{eff}}+\alpha\m\times\frac{d\m}{dt}+{\boldsymbol{\tau}}_{\text{d}}+{\boldsymbol{\tau}}_{\text{f}}.
\end{equation}
The four terms on the right side of the above equation describe the gyromagnetic precession, the dissipation originating from the nonzero Gilbert damping, and the spin torques induced by the driving current, respectively. It should be noted that we focus on the damping-like spin torque ${\boldsymbol{\tau}}_{\text{d}}=\gamma u_{\text{d}}\m\times(\p\times\m)$ and ignore the field-like torque ${\boldsymbol{\tau}}_{\text{f}}=\gamma u_{\text{f}}\p\times\m$ in the main text since the latter is regarded as a weak term in the considered material system without a  sufficiently strong interfacial Rashba effect. However, without loss of generality, the effect of the field-like torque on the oscillation has also been investigated and is shown in Supplementary Figure 8. It is found that a large field-like torque can decrease or increase the oscillation frequency, depending on the sign of $u_{\text{f}}/u_{\text{d}}$. Here, $\m$ denotes the local normalized magnetization, $\gamma$ is the gyromagnetic ratio, $\alpha$ is the damping constant and $\p$ is the spin polarization direction of the spin current. $u_{\text{d}}$ and $u_{\text{f}}$ are proportional to the current density $j$, and describe the magnitudes of the damping-like torque and field-like torque, respectively. $\H_{\text{eff}}$ represents the effective magnetic field associated with the total energy of the system,
\begin{equation}
\label{eq:energy}
E=\int{[A(\boldsymbol{{\nabla}\m})^2-Km^2_{z}+\varepsilon_{\text{d}}+\varepsilon_{\text{DMI}}}]dV,
\end{equation}
where $A$ is the Heisenberg exchange stiffness, $K$ is the PMA constant and $\varepsilon_{\text{d}}$ denotes the demagnetization energy density. The last term arises from the asymmetric exchange interaction DMI, which is related to the spin configuration and the chirality of magnetic structures. For example, the Bloch-type skyrmion is stabilized by the bulk DMI $\varepsilon_{\text{DMI}}=D\m\cdot (\boldsymbol{{\nabla}}\times\m)$, the N{\'e}el-type skyrmion is formed in the thin film with the interfacial DMI $\varepsilon_{\text{DMI}}=D[(\m\cdot\boldsymbol{{\nabla}})m_z-m_z(\boldsymbol{{\nabla}}\cdot\m)]$, while for the magnetic system with crystallographic class $D_{2d}$ that tends to stabilize an antiskyrmion, the DMI energy density is given by $\varepsilon_{\text{DMI}}=D\m\cdot(\partial_x \m\times\hat{x}-\partial_y \m\times\hat{y})$~\cite{Cortes-Ortuno_NJP2018}. Meanwhile, the sign of DMI constant $D$ designates the chirality of skyrmion-like structures, i.e., the left-handedness or right-handedness. In this work, we focus on the N{\'e}el skyrmion, skyrmionium and antiskyrmion, which are realized by considering different DMIs, either the crystallographic class or the strength. 

\noindent
\textbf{Magnetic parameters.}
The primary magnetic parameters of Pt/Co magnetic film are adopted~\cite{Sampaio_NATTECH2013}: saturation magnetization $M_s=580$ kA$\cdot$m$^{-1}$, exchange stiffness $A=15$ pJ$\cdot$m$^{-1}$, PMA constant $K=0.8$ MJ$\cdot$m$^{-3}$ in the clean magnetic area without pinning sites. The DMI constant is set to be $3.5$ mJ$\cdot$m$^{-2}$, at which both the skyrmion and the skyrmionium are the metastable states corresponding to the minimal values in the energy profile~\cite{Rohart_PRB2013}. It should be noted that the oscillation discussed in our work is not limited to this set of parameters, but suitable for a wide range of magnetic parameters. It can be extended to most of the existing materials that support skyrmions. Supplementary Figures 9 and 10 also provide stability diagrams of the elongated skyrmion for different combinations of magnetic parameters. To study the skyrmion oscillation, the rectangular mesh of $1.0\times 1.0\times 0.4$ nm$^3$ is used to discrete a ferromagnetic film with the dimension of $200\times 100\times 0.4$ nm$^3$  and the Gilbert damping coefficient is 0.1. However, for the discussion about the propagation of spin waves in an elongated skyrmion, the model size is expanded to $400\times 200\times 1$ nm$^3$ and the mesh size becomes $1\times 1\times 1$ nm$^3$. The Gilbert damping constant is reduced to 0.05 to suppress the dissipation of spin waves in the simulation. 
%

\vbox{}
\section{\sffamily Data availability}

\noindent
The data that support the findings of this study are available from the corresponding author upon
reasonable request.

\vbox{}
\section{\sffamily Code availability}

\noindent
The micromagnetic simulation software OOMMF used in this work is open-source and can be accessed freely at http://math.nist.gov/oommf. 
%

\bibliography{communphys} 

\vbox{}
\section{\sffamily Acknowledgments}
This research is supported by the Guangdong Basic and Applied Basic Research Foundation (2021B1515120047), the Guangdong Special Support Project (2019BT02X030), the Shenzhen Fundamental Research Fund (Grant No. JCYJ20210324120213037), the Shenzhen Peacock Group Plan (KQTD20180413181702403), the Pearl River Recruitment Program of Talents (2017GC010293) and the National Natural Science Foundation of China (11974298, 61961136006).

\vbox{}
\section{\sffamily Author contributions}
Y.Z. coordinated the project. X.L. performed the numerical simulations and theoretical calculations with help from L.S. X.L. drafted the manuscript and revised it with input from L.S., X.X. and Y.Z. All authors discussed the results and reviewed the manuscript.

\vbox{}
\section{\sffamily Competing interests}
The authors declare no competing interests.

\vbox{}
\section{\sffamily Supplementary information}
See the supplementary information for details on the generation of an elongated skyrmion, the conversion between an ordinary skyrmion and an elongated skyrmion, effects of the magnetic parameters on the elongated skyrmion, the stability diagram of elongated skyrmion for different magnetic parameters, effects of the field-like torque on the oscillation frequency, the schematics of the proposed device as a magnonic waveguide and the proposal of a logic gate based on the propagation of spin waves.

\end{document}